\documentclass[aps]{revtex4-1}
\newcommand{\beq}{\begin{equation}}
\newcommand{\eeq}{\end{equation}}
\newcommand{\bea}{\begin{eqnarray}}
\newcommand{\eea}{\end{eqnarray}}
\newcommand{\uvec}[1]{{\bf \hat{#1}}}
\newcommand{\eq}[1]{{eq. (\ref{#1})}}

\usepackage{graphicx}

\begin{document}
\title{Spin-Transfer-Torque Driven Magneto-Logic Gates Using Nano Spin-Valve Pillars}

\author{C. Sanid and S. Murugesh}
\email{murugesh@iist.ac.in}
\affiliation{ 
Department of Physics, Indian Institute of Space Science and Technology,\\
Trivandrum - 695~547, India.
}%

\begin{abstract}
{We propose  model magneto-logic NOR and NAND gates using a spin valve 
pillar, wherein the logical operation is induced by spin-polarized currents 
which also form the logical inputs. The operation is 
facilitated by the simultaneous presence of a constant {\it controlling} 
magnetic field. The same spin-valve assembly can also be 
used as a magnetic memory unit. We identify
regions in the parameter space of the system where the logical operations can 
be effectively performed. The proposed gates retain the non-volatility of 
a magnetic random access memory\,(MRAM). 
We verify the functioning of the gate by numerically simulating its
dynamics, governed by the appropriate Landau-Lifshitz-Gilbert equation  
with the spin-transfer torque term. The flipping time for the logical states 
is estimated to be within nano seconds. }
\end{abstract}

\maketitle

\section{Introduction}
Following the discovery of the celebrated giant magneto-resistance\,(GMR) 
effect and the development of 
spin-valve structures, a reciprocal phenomenon of GMR, a torque on the
spin magnetization induced by spin polarized currents, was independently 
predicted by Slonczewski and Berger\cite{slonc:1996,berger:1996}. 
In this 
spin-transfer torque\,(STT) effect a spin polarized current flowing perpendicular
to a thin ferromagnet generates a  
torque strong enough to reorient its magnetization. 
Information coded in the form of macrospin of the magnetic layer, considered
as a monodomain, is thus amenable to manipulation using spin-polarized currents
\cite{stiles:2006,wolf:2006}. The 
extensive theoretical and experimental studies on spin valve geometries that
followed brought into light two 
especially important phenomena relevant to magnetic storage technology and 
spintronics - current induced magnetization switching and self-sustained 
microwave oscillations in the nanopillar devices\cite{myers:1999,gro:2001,kiselev:2003,rippard:2004,berkov:2008}. 

The aspect of {\it non-volatility}, fundamentally inherent in the system,
and the significant reduction in power consumption have prompted the
development of spin-valves as memory devices.  
The earlier proposals, however, were based on a field induced 
magnetic switching(FIMS) approach for writing data, which uses two orthogonal 
pulses of magnetic filed to achieve writing.
Magnetic random access memory\,(MRAM) models based on current induced magnetic switching (CIMS), 
wherein STT phenomenon forms the core, have since been proposed. 
Apart from the more obvious application as plain memory storage devices, 
spin valve based magneto-logic devices have also been attempted in the recent 
past. FIMS based field programmable logic gates using GMR elements were proposed
by Hassoun {\it et al.}\cite{black:1997}, wherein the type of the logical 
operation to be performed can be altered by additional fields.
Further models have also been suggested where the logical state of the GMR 
unit is manipulated using FIMS \cite{richter:2002,koch:2003,ney:2005,wang:2005,chao:2007}. Similar programmable models 
based on spin valve magneto-logic
devices are also known in literature\cite{zhao:2006,dery:2007,buford:2011}. 
These
later models, based on CIMS, involve additional spin-valve elements that 
together form a single logical unit, or more than one current carrying plate 
capable of generating fields in orthogonal directions. 
Besides, in these
models, bi-polar
currents were crucial in writing or manipulating data. Invariably, this
requires a more complex architecture than is required for a simple magnetic 
memory unit. 

In this paper we propose alternative magneto-logic NAND and NOR gate models, 
wherein the logical operation is performed through CIMS in the presence of
a controlling field. Apart from the simplicity in the architecture, the 
models also carry the advantage that they can be 
used as plain memory elements in a MRAM.
They consist of a single spin-valve pillar and no additional elements,
than those required for its functioning as a memory unit, are required to
enhance its role as a logical gate. In 
the proposed models we use STT for writing, while the magnetic field is held 
constant in magnitude (positive for NOR and negative for NAND gate) and 
required only during the logical operation. Thus the applied field 
acts as a control switch for the gates. The gates are also non-volatile, 
as naturally expected in a magneto-logic device. 

\section{Spin-Valve Pillar Geometry and the Governing Landau-Lifshitz-Gilbert 
Equation}

The system under consideration is a regular spin valve primarily consisting of 
a conducting layer sandwiched between two ferromagnetic layers, one pinned 
and the 
other free, with the magnetization in the pinned layer parallel to the plane
of the free layer - a geometry that is well studied\cite{kiselev:2003,albert:2000,mangin:2006}. 
Further, the free layer is also subject to a constant Oersted 
field, by a conducting plate carrying current. The dynamics of the macrospin 
magnetization of the free layer is 
governed by the Landau-Lifshitz-Gilbert\,(LLG) equation with the STT term, whose dimensionless form 
is given by\cite{berkov:2008,bert:2005}
\begin{equation}\label{llg}
\frac{\partial \textbf{m}}{\partial t}-\alpha\textbf{m}\times\frac{\partial \textbf{m}}{\partial t} = - \textbf{m} \times \textbf{H}_{eff},
\end{equation}
where
\[
\textbf{H}_{eff} = \left( \textbf{h}_{eff}-\beta\frac{\textbf{m}\times\textbf{e}_{p}}{1+c_{p}\textbf{m}\cdot\textbf{e}_{p}}\right).
\]
The free-layer magnetization $\textbf{m}$ and the effective field 
$\textbf{h}_{eff}$ are normalized by the saturation magnetization $M_s$. Time is
measured in units of $(\gamma M_s)^{-1}$, where $\gamma$ is the gyromagnetic
ratio (for Co layers, this implies
time scales in the order of picoseconds). The constant $\alpha$ is 
the damping factor and unit vector $\textbf{e}_{p}$ is the direction of pinning 
($\uvec{x}$ in our case, and in plane). The other constant 
$c_{p}\,({1}/{3}\le c_{p}\le1)$ is a function of degree of spin polarization 
$P\,(0\le P\le1)$:
\begin{equation}\label{cp}
  c_{p}=\frac{(1+P)^{3}}{3(1+P)^{3}-16P^{{3}/{2}}}
\end{equation}
In the numerical calculations that follow we have used the typical value 
of $P=0.3$. The phase diagrams, to be discussed in the next section, do 
exhibit minor variations with change in the value of $P$, but do not alter 
our results much. For, as can be seen from \eq{cp}, 
$c_{p}$ is a small number compared to $1$ for all realistic values of $P$.
The parameter $\beta$ is proportional to the spin 
current density (typically of the order of $10^{-2}$ for Co layers, with
current densities $\sim 10^{8}$~A/cm$^2$). 
The effective field is given by 
\[
\textbf{h}_{eff} = h_{ax}\uvec{x}-(D_{x}m_{x}\uvec{x}+D_{y}m_{y}\uvec{y}+D_{z}m_{z}
\uvec{z}),
\]
where $h_{ax}\uvec{x}$ is the external field and $D_{i}$s$(i=x,y,z)$ are constants 
that reflect the crystal shape and anisotropy effects. Particularly, we chose
our film such that the anisotropy is in-plane, and also lies along the $x$-axis.
The plane of the free layer 
is chosen to be the $x-y$ plane. With this choice $D_is$ are such that  
$D_{x}<D_{y}<D_{z}$, making $\uvec{x}$ the free-layer easy axis.

\section{Geometry of Fixed Points and Magneto-Logic Gates}
\begin{figure}[h]\label{para}
\begin{center}
\includegraphics[width=0.8\linewidth]{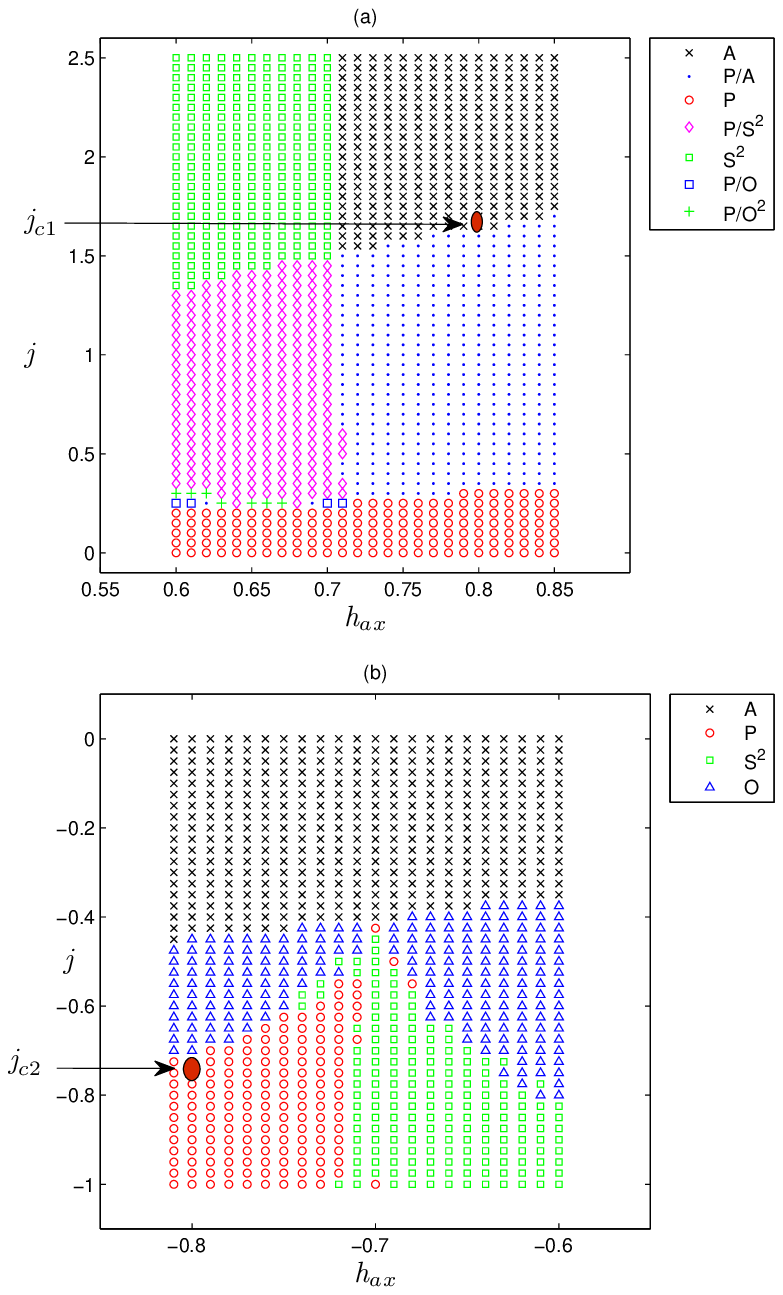}
\caption{Phase diagram in the $h_{ax}-j$ space, in regions relevant for the 
(a) NOR and (b) NAND gates. The system displays limit cycles(O), symmetric
out-of-plane limit cycles\,(O$^2$), stable fixed points parallel to 
$\uvec{x}$(P)
or $-\uvec{x}$(A), and symmetric out-of-plane stable fixed points (S$^2$). The
critical value of the current and the field used for our models ($j_{c1}$ 
and $j_{c2}$) are circled in the two figures.}
\end{center}
\end{figure}
For our choice of geometry, described in the previous
section, the magnetization in the free layer exhibits a variety of dynamics 
in different regions of the $h_{ax}-j\,(\equiv\beta/\alpha)$ parameter space -
such as in-plane limit cycles (O) and symmetric out-of-plane limit cycles 
(O$^2$), and stable fixed points parallel to $\uvec{x}$(P), parallel to 
$-\uvec{x}$(A) and symmetric out-of-plane stable fixed points 
(S$^2$)\cite{bert:2005,bert:2008}. In Fig. 1 we show two specific
ranges where the models
we propose can perform the desired logical operations. The type of dynamics 
in the different regions of the parameter space is identified here by 
numerically simulating the LLG equation \eq{llg}. These results clearly agree
with those obtained analytically in \cite{bert:2005,bert:2008}. 

\begin{figure}[h]\label{fp}
\begin{center}
\includegraphics[width=0.75\linewidth]{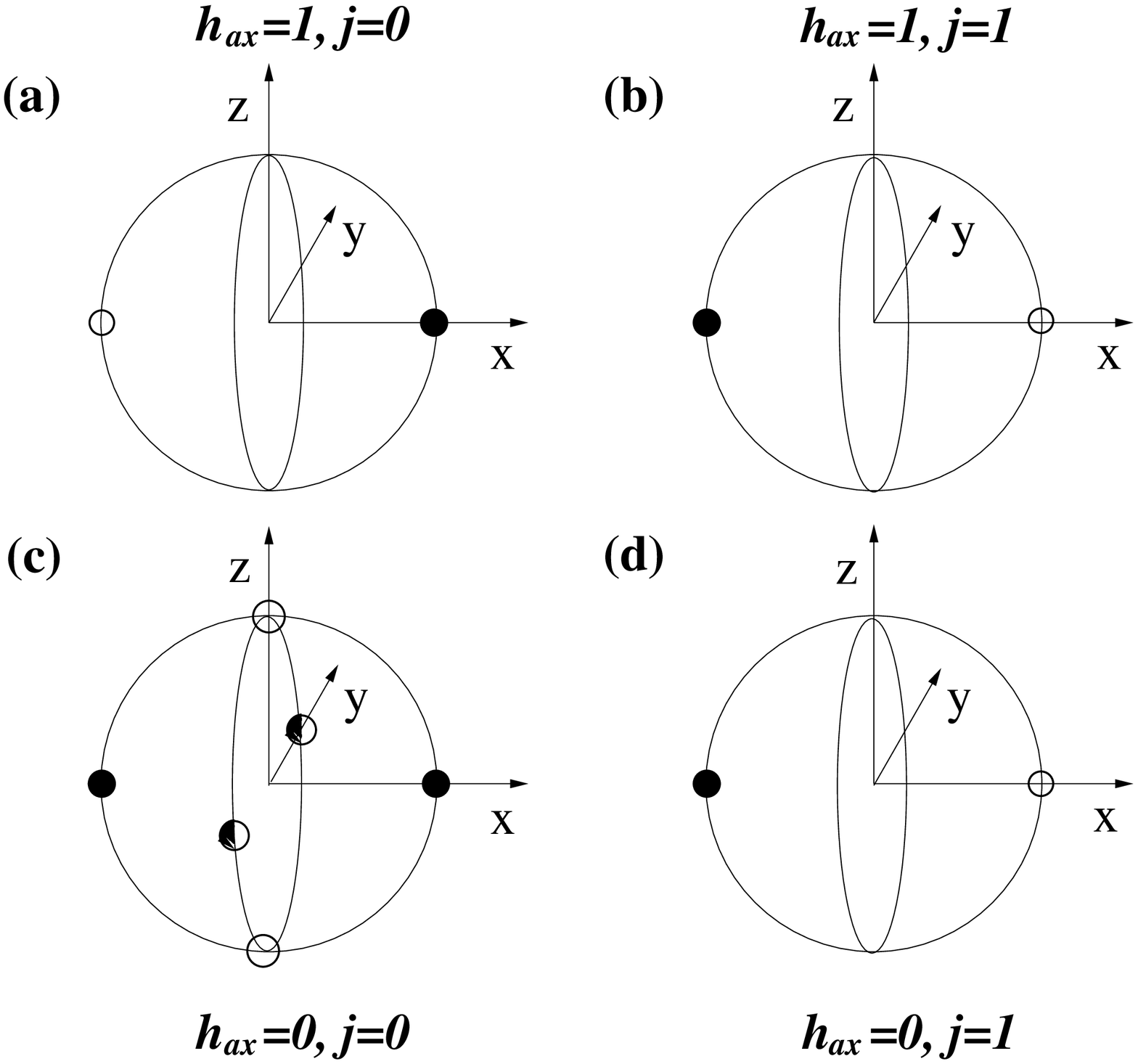}
\caption{Fixed points for four different cases. For convenience, we have 
indicated the
fixed value of the applied field we have used through out,
satisfying the condition $|h_{ax}|>D_z-D_x$, as $h_{ax}=$1. Similarly the value
for $j\,(>j_{c1})$, is
indicated by $j=1$.  (a) $h_{ax}=1,~j=0$, 
(b) $h_{ax}=1,~j=1$, (c) $h_{ax}=0,~j=0$, and (d) $h_{ax}=0,~j=1$.  
Stable fixed points are indicated by filled dots, and unstable fixed points
by unfilled dots. For $h_{ax}=0=j$, there arise six fixed points, two of which
are saddles indicated by half filled dots, and both $\pm\uvec{x}$ are
stable fixed points. }
\end{center}
\end{figure}

\subsection{Logic NOR gate}
For the logical NOR gate, we shall choose the applied field (whenever non-zero)
to be positive and $|h_{ax}|>D_z-D_x$. 
For this choice, there can at best
be only one stable fixed point, lying along either $\pm\uvec{x}$ 
directions  depending on the values of $h_{ax}$ and $j$.
For a given set of values of the system parameters, $D_i$s and $\alpha$, fixed
points corresponding to four scenarios of special interest to us in designing 
our NOR gate are shown in Fig. 2. 
When $j$ is held below a certain threshold value, 
and $|h_{ax}|>D_z-D_x$, 
${\bf m}=\uvec{x}$ is the only stable fixed 
point, while $-\uvec{x}$ is unstable. For $j$ beyond a certain upper threshold 
value $j_{c1}$, with $h_{ax}$ held at the same value, the situation reverses, 
with $\uvec{x}$ becoming unstable and
$-\uvec{x}$ becoming the stable point. 
When $h_{ax}=0=j$, both $\pm\uvec{x}$
become stable on account of the anisotropy field along the ${\bf x}$ axis. 
Finally, when $h_{ax}$ is held at zero, but $j>j_{c1}$, the scenario in Fig. 2(b)
repeats, with $-\uvec{x}$ stable and $\uvec{x}$ unstable.

\begin{figure}[h]\label{logic}
\begin{center}
\includegraphics[width=1\linewidth]{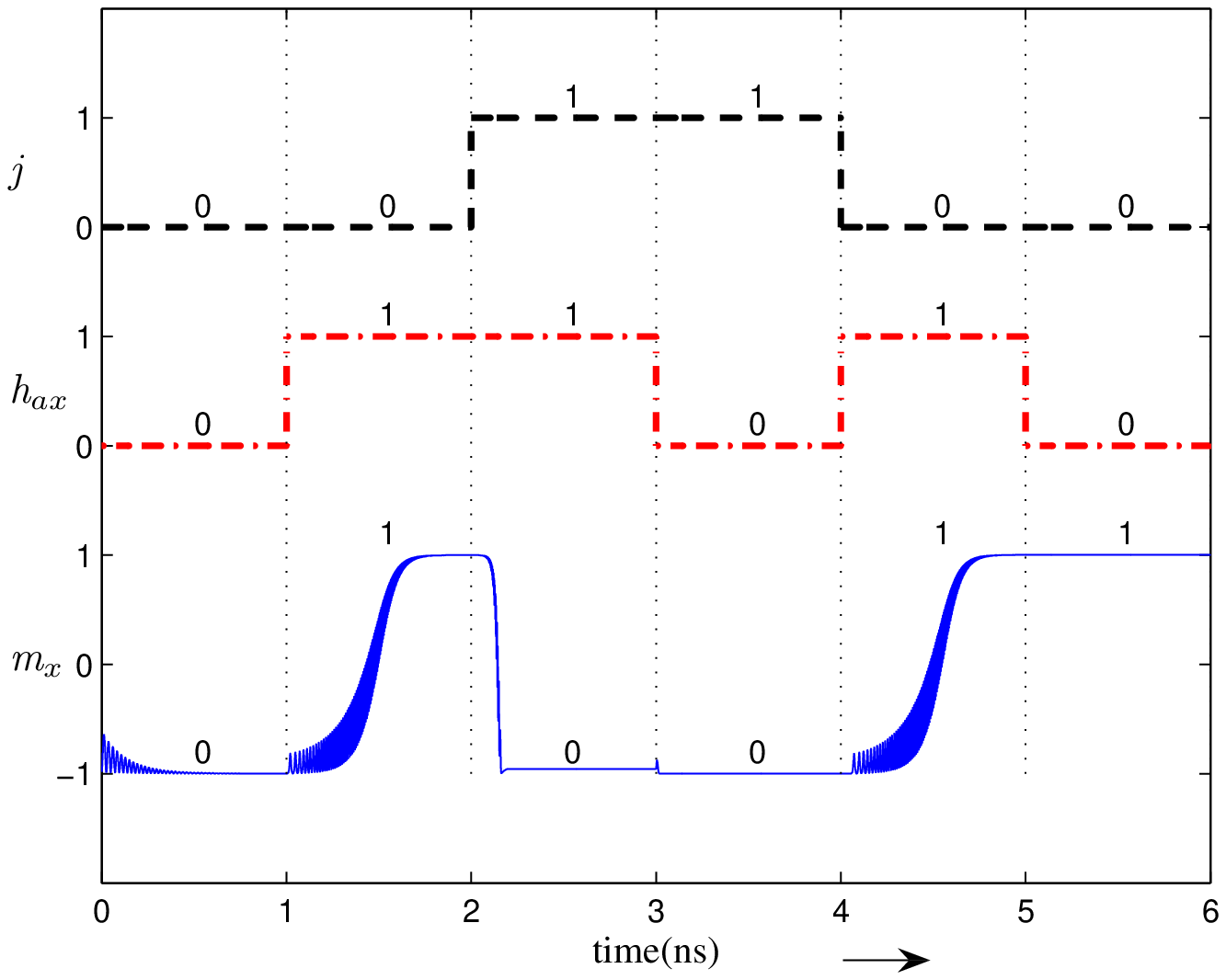}
\caption{Time evolution of $m_x$ (bottom) as the applied field $h_{ax}$ 
(middle) and $j$ (top)
are flipped through various combinations, with the interpreted logical state.
The initial orientation of ${\bf m}$
is chosen arbitrarily. For the case $h_{ax}=0=j$, both $\pm\uvec{x}$ are
stable fixed points, and the magnetization relaxes to the nearest of the
two directions-$m_x=-1$ initially, and $m_x=+1$ finally.}
\end{center}
\end{figure}
A numerical simulation of the governing LLG equation with the STT term,
\eq{llg}, shows the expected magnetization switching in conformity with 
Fig. 2. 
We choose the system parameters $\alpha=0.01$, $D_x=-0.034$, $D_y=0$,
and $D_z=0.68$ (as in ref. \cite{kiselev:2003}). Taking the value of saturation 
magnetization, $M_s$,
to be that of Co ($1.4\times~10^6$~A/m), it effectively implies a time
scale of $3.2$~ps. The switching time due to the spin-current is roughly 
0.2~ns, while that due to the magnetic field is slower, at nearly 0.7~ns,
accompanied by a ringing effect. 
This delay and ringing effect are well understood to be due to the fact that, 
even with $\alpha=0$, a spin-transfer-torque
leads to both precession and dissipation whereas a 
magnetic field alone 
can only cause a precession of magnetization vector about the applied 
field\cite{sm:2009}. Field induced switching 
is thus exclusively due to the damping factor, leading to a longer switching
time, consequently. A longer switching time invariably implies more 
precession meanwhile, causing the ringing effect. 
In Figure 3, we show the dynamics of the $x$ component 
of the normalized magnetization vector ${\bf m}$ as the field and current are
switched through various possible combinations. The current density used 
is of the order of $10^8$~A/cm$^2$, and the field $h_{ax}$ is of the 
order of $10^6$~A/m. Such a magnitude for the applied field, although
frequently used (see, for instance ref. \cite{kiselev:2003}), is substantially
high for real world applications. Magnetic tunnel junctions (MTJs)  have
proved themselves to be more worthwhile candidates as MRAMs, with their 
operability at much lower spin-current and field amplitudes, and higher
ferromagnetic to anti-ferromagnetic current 
ratios\cite{alan:2009,parkin:2003,daughton:1997}. 
Although the STT phenomenon in MTJs and that in spin-valve pillars
display several qualitative similarities, MTJs are hampered
by the lack of an appropriate mathematical model to describe their dynamics. 
We believe results presented in this paper will be of relevance in MTJs 
too and may possibly be reproduced. Our numerical simulations show that the 
model presented is robust with respect to errors that may creep in through two
of the system parameters - variations in the degree of polarization, and in 
plane anisotropy fields in the form of $D_x$. We have varied these values 
upto 10\% and yet noticed no percievable difference in the phase diagram. 
The chosen values of $j_i's$ ($0.6j_{c1}$) provides enough
room for errors arising out of fluctuations. Further, we recall that as long 
as the condition $|h_{ax}|>D_z-D_x$ is satisfied we have the two desired fixed 
points, enabling the required logical operation.

We make use of the first three scenarios (Figs. 2(a)-2(c)) to construct the 
universal NOR gate, 
which retains the non-volatility of spin based memory devices. Let $j_1$ 
and $j_2$ be currents that form inputs to the logic gate, and 
each take either of the two values - {\it zero}, or some value $j$ little 
over $j_{c1}$. We shall identify these values of the current with the 
logical input states 0 and 1, respectively. Both currents $j_1$ and $j_2$
are fed together into the spin-valve from the pinned layer end. 
The field $h_{ax}$ is held fixed throughout
the logical operation (represented henceforth simply 
as $h_{ax}=$1), 
and acting as a controlling 
field. When the currents $j_{1,2}$ are both zero, the magnetization ${\bf m}$
orients itself along $\uvec{x}$, the only stable fixed point. 
This corresponds to the low resistance state, being 
{\it parallel} to the pinned layer magnetization, which we read as the logical 
state 1. When either, or both, of the currents
$j_{1,2}$ is greater than $j_{c1}$, the torque is sufficient enough to flip the
spin ${\bf m}$ from any direction to the new stable fixed point $-\uvec{x}$ 
(the high resistance {\it anti-parallel} state 0). 
The following truth table of the NOR gate is thus obtained (see Table 1).
When the field $h_{ax}$ and the currents $j_{1,2}$ are all switched off, 
both $\pm\uvec{x}$ are equally good stable fixed points
due to the anisotropy field along the ${\bf x}$ axis.  
Prior value of magnetization is therefore
retained, and the gate carries the non-volatility of the MRAM. 
\begin{center}
\begin{table}[h]\label{truth}
	\begin{tabular*}{0.75\columnwidth}{@{\extracolsep{\fill}} | c | c | c | c |}
	\hline
	$h_{ax}$ & $j_{1}$ & $j_{2}$ & ${\bf m}$~(logical state)\\ 
	\hline
	1 & 0 & 0 & $\uvec{x}$ (1)\\ 
	\hline
	1 & 1 & 0 & $-\uvec{x}$ (0)\\ 
	\hline
	1 & 0 & 1 & $-\uvec{x}$ (0)\\ 
	\hline
	1 & 1 & 1 & $-\uvec{x}$ (0)\\ 
	\hline
	\end{tabular*}
\caption{The truth table for NOR gate. The applied field
is always held constant through out the operation ($|h_{ax}|>D_z-D_x$) 
indicated by $h_{ax}=1$. The currents $j_{1,2}$ take either a value greater 
than $j_{c1}$, indicated as
the logical input 1, or zero taken as input 0.}
\end{table}
\end{center}

\begin{figure}[h]\label{fp2}
\begin{center}
\includegraphics[width=1\linewidth]{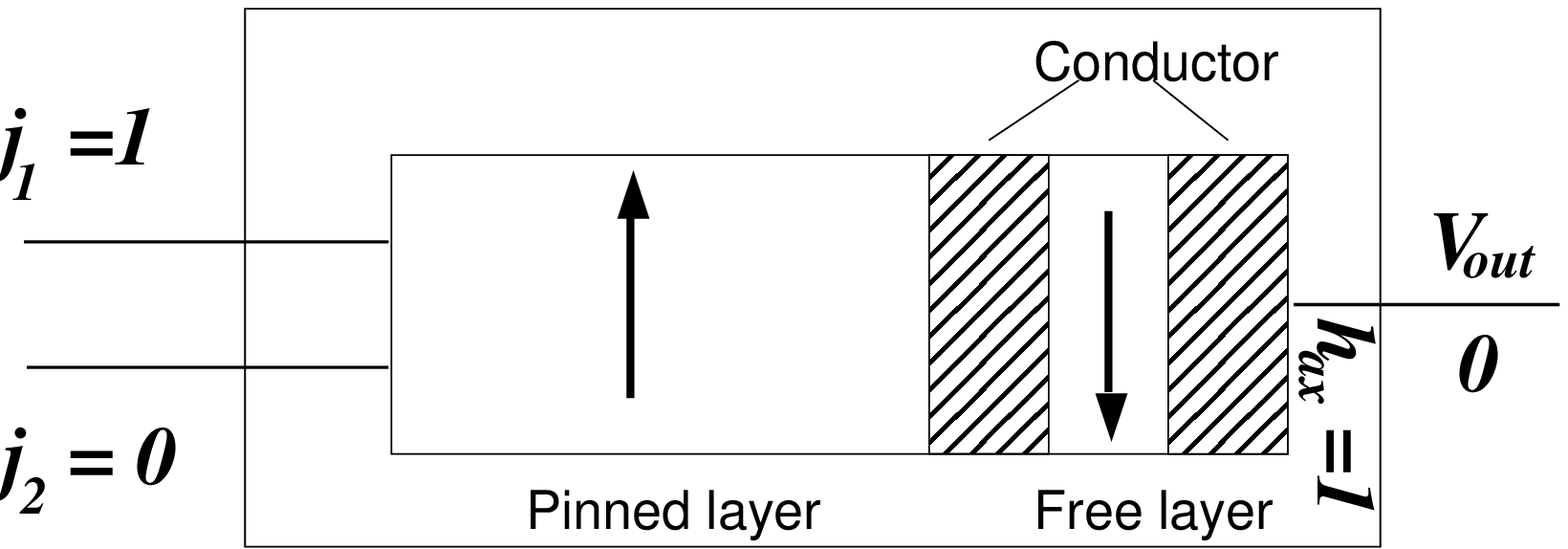}
\caption{A schematic diagram of the NOR gate, with the relevant 
portion of the spin-valve pillar, and a specific set of values for
the input currents $j_i's$ and the control field $h_{ax}$. The logical
output is interpreted from the value of the potential $V_{out}$, either high 
(state 1) or low (state 0). }
\end{center}
\end{figure}

The nature
of fixed points depicted in Figs. 2(a), 2(c), and 2(d), show that
the same valve assembly can also be used as a plain memory device. To this
end we shall use a single current input, $j$, to the spin-valve 
as opposed to the two inputs for the gate assembly. 
{\it Writing} the data bit 1 is then
enabled with a applied field
$h_{ax}=1$ and current $j=0$. Similarly the bit 0 is written when
$h_{ax}=0$ and $j=1$. The two stable fixed points, as shown in Fig. 2(c),
then ensure that the magnetization, or {\it data}, is retained in the absence 
of both the current and field, preserving non-volatility. A schematic 
representation of the logical NOR gate for a choice of input currents,
and with control field $h_{ax}=1$, is shown in Fig. 4.

\begin{figure}[h]\label{fp2}
\begin{center}
\includegraphics[width=0.75\linewidth]{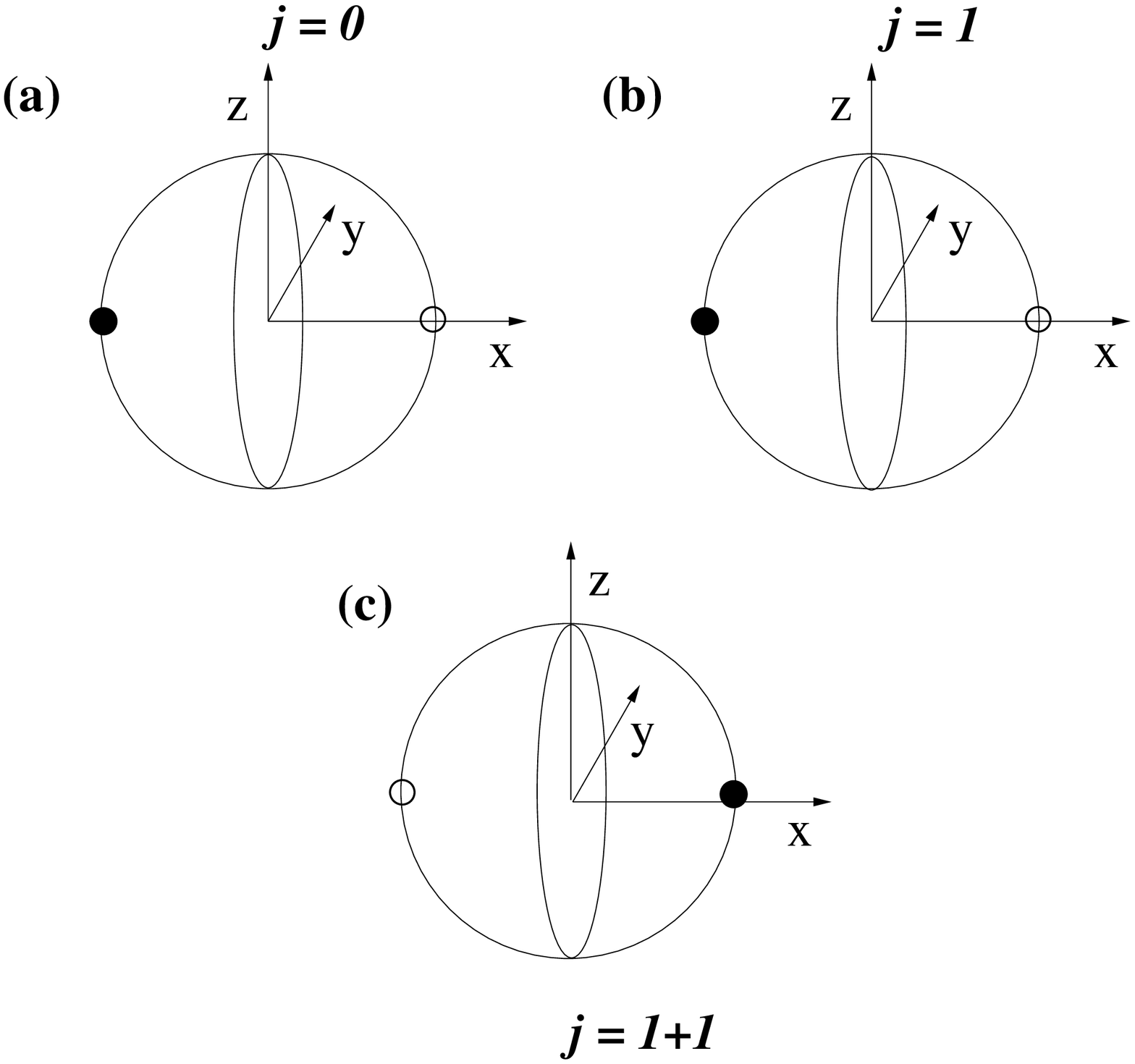}
\caption{Fixed points for three different values of the current $j$,
(a) $j=0$, (b) $j=1\,(0.6j_{c2})$ and (c) $j=1+1\,(1.2j_{c2})$. The applied 
field $h_{ax}$ is the same, and is negative with $|h_{ax}|>D_z-D_x$. 
When both field and current are zero, the fixed points are the same as 
in Fig. 1(c).}
\end{center}
\end{figure}

\subsection{Logic NAND gate}
We now look at the fixed points corresponding to
another region of the $h_{ax} - j$ parameter space [Fig. 1(b)]. The
applied field $h_{ax}$ is chosen to be {\it negative} (again, whenever 
non-zero), while still satisfying the earlier condition that $|h_{ax}|>D_z-D_x$,
and the current $j$ assumes
either of the three values, {\it zero}, $0.6j_{c2}$ or  $1.2j_{c2}$ [where
$j_{c2}$ is indicated in Fig. 1(b)]. Notice that $j_{c2}$ is {\it negative},
implying a current sent in the opposite direction along the pillar. 
The fixed points corresponding
to different combinations of $h_{ax}$ and $j$ are shown in Fig. 5. 
We shall denote the above mentioned negative value of the magnetic
field as $h_{ax} =-1$. For the NAND gate we shall take the current 
value $j=0$, and $j=0.6j_{c2}$ as the logical inputs 0 and  1, 
respectively. In the absence of both current and field, the stable fixed points
are $\pm\uvec{x}$, as in Fig. 2(c). When the field $h_{ax}=-1$ and the 
current is either 0 or 1, $\uvec{m}=-\uvec{x}$ is the only stable 
fixed point while $\uvec{m}=\uvec{x}$ becomes unstable. When the current
value $j=1.2j_{c2}$, however, the situation reverses, with $\uvec{x}$ 
becoming stable, and $-\uvec{x}$ unstable.  A numerical simulation, analogous
to Figure 3, for these new values of $h_{ax}$ and $j$ is shown in Figure 5,
with results as expected. 
\begin{figure}[h]\label{nand}
\begin{center}
\includegraphics[width=1\linewidth]{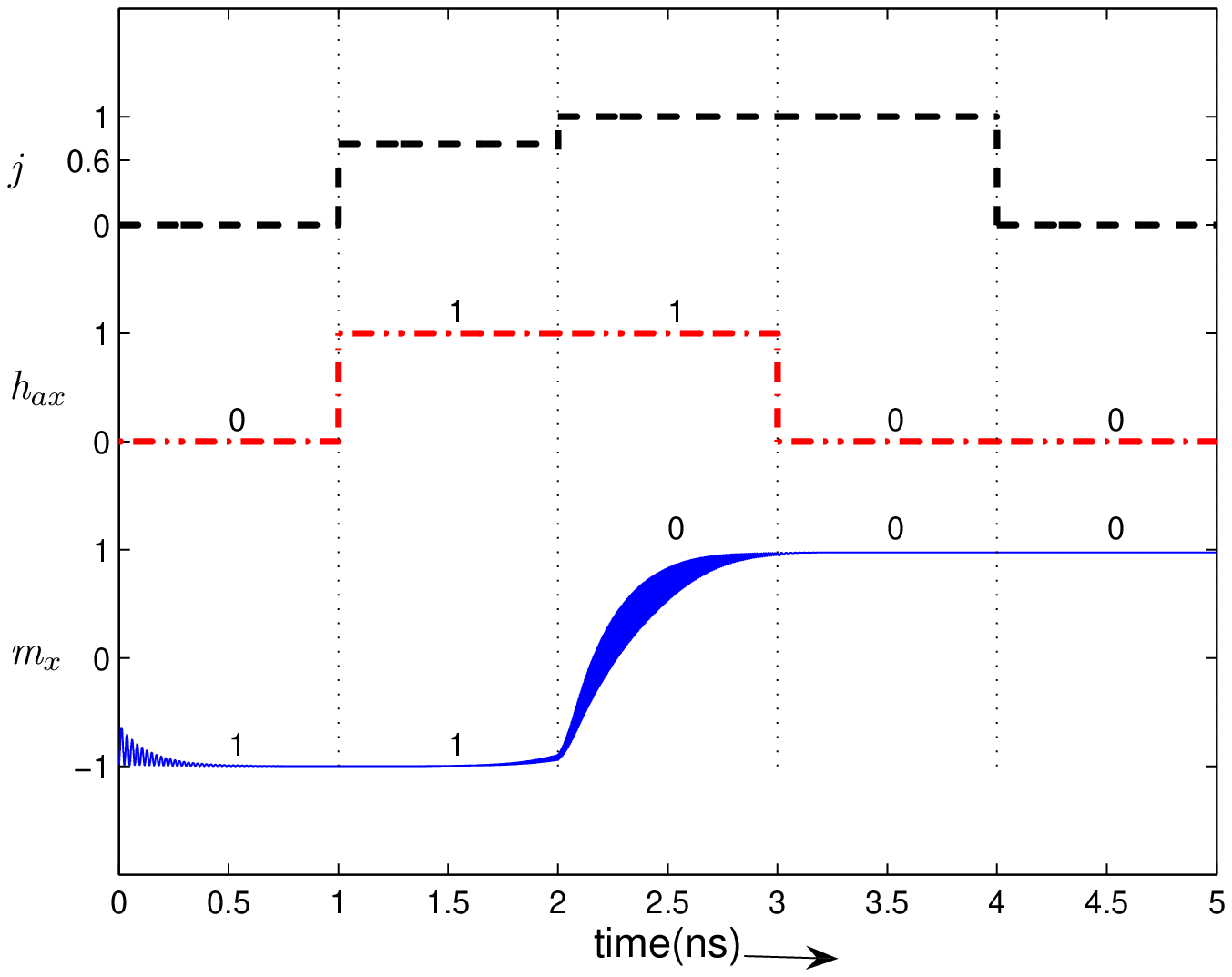}
\caption{Time evolution of $m_x$ (bottom) as the applied field $h_{ax}$ 
(middle) and $j$ (top)
are flipped through various combinations, relevant to the NAND gate.
The interpreted logical state is indicated over the respective $m_x$ values.
For the parameter values chosen, the switching time is within 1~ns. }
\end{center}
\end{figure}

As in the case of the NOR gate, let
$j_1$ and $j_2$ be the currents fed together, and each take values 0 or 1
(now corresponding to negative currents). The magnetic field is held
constant at $h_{ax}=-1$ all along the logical operation. 
For the logical NAND gate we adopt the opposite convention, interpreting 
the high-resistance state ($\uvec{m} = -\uvec{x}$) as the logical state 1,
and the low-resistance state as 0. The following
truth table of the NAND operation is thus realized (Table 2). 
As both $\pm\uvec{x}$ are stable fixed points in the absence of current
and magnetic field [Fig. 2(c)], non-volatility is ensured.
\begin{center}
\begin{table}[h]\label{truth2}
	\begin{tabular*}{0.75\columnwidth}{@{\extracolsep{\fill}} | c | c | c | c |}
	\hline
	$h_{ax}$ & $j_{1}$ & $j_{2}$ & ${\bf m}$~(logical state)\\ 
	\hline
	-1 & 0 & 0 & $-\uvec{x}$ (1)\\ 
	\hline
	-1 & 1 & 0 & $-\uvec{x}$ (1)\\ 
	\hline
	-1 & 0 & 1 & $-\uvec{x}$ (1)\\ 
	\hline
	-1 & 1 & 1 & $\uvec{x}$ (0)\\ 
	\hline
	\end{tabular*}
\caption{The truth table for NAND gate. As earlier, the applied field
is always held constant through out the operation, though negative. The 
currents $j_{1,2}$ take either of the two values
$0.6j_{c2}$ - the logical input 1, or zero taken as input 0.}
\end{table}
\end{center}


\section{Summary}
In summary, we have proposed spin-valve based magneto-logic NOR and
NAND gate assemblies, which render
themselves to the dual role of universal gate and a magnetic memory.  
A constant applied magnetic field parallel to the pinned layer magnetization
acts as a control for the logic gate operation, while spin-currents are
fed in as the logical inputs. The same pillar geometry is used for both
the NOR and NAND gates, and also doubles as a magnetic memory device.

\end{document}